\documentclass[reprint, amsmath,amssymb,aps,prd]{revtex4-1}
\usepackage{subfigure}
\usepackage{graphicx}
\usepackage{dcolumn}

\usepackage{bm}
\usepackage{xcolor}
\usepackage[colorlinks,urlcolor=aa,linkcolor=blue,anchorcolor=aa,citecolor=aa]{hyperref}
\usepackage[mathlines]{lineno}
\usepackage{multirow}
\usepackage{enumerate}
\usepackage{amsmath}

\usepackage{lineno}

\usepackage{mathtools}
\usepackage{overpic}
\usepackage{epsfig}
\usepackage{rotating}
\usepackage{longtable}
\usepackage{color}
\usepackage{siunitx}
\usepackage{diagbox}

\definecolor{aa}{RGB}{0,0,139}

\newcommand{\BR}{\mathcal{B}}

\newcommand{\gev}{\rm GeV}

\newcommand{\gevcs}{{\rm GeV}/c^2}
\newcommand{\mev}{\rm MeV}
\newcommand{\mevcs}{{\rm MeV}/c^2}
\newcommand{\kev}{\rm keV}

\newcommand{\ks}{K_S^0}

\newcommand{\cm}{\si{\centi\metre}}
\newcommand{\ns}{\si{\nano\second}}
\newcommand{\ps}{\si{\pico\second}}

\newcommand{\eff}{\varepsilon}

\newcommand{\piz}{\pi^0}

\newcommand{\etacp}{\eta_c(2S)}

\newcommand{\psp}{\psi(3686)}
\newcommand{\jpsi}{J/\psi}
\newcommand{\chicz}{\chi_{c0}}
\newcommand{\chico}{\chi_{c1}}
\newcommand{\chict}{\chi_{c2}}

\newcommand{\EE}{e^+e^-}
\newcommand{\MM}{\mu^+\mu^-}

\newcommand{\GG}{\gamma\gamma}
\newcommand{\pp}{\pi^+\pi^-}
\newcommand{\ppp}{\pi^+\pi^-\pi^0}
\newcommand{\kk}{K^+K^-}
\newcommand{\kkp}{\kk\piz}
\newcommand{\kskp}{\ks K^{\pm} \pi^{\mp}}

\newcommand{\kkpII}{K \bar{K} \pi}

\newcommand{\ptge}{\psp\to\gamma\etacp}

\newcommand{\etkkp}{\etacp\to \kkp}

\newcommand{\gfake}{\gamma_{\rm fake}}
\newcommand{\gisr}{\gamma_{\rm ISR}}
\newcommand{\gfsr}{\gamma_{\rm FSR}}

\newcommand{\reduline}{\bgroup\markoverwith
{\textcolor{red}{\rule[0.5ex]{2pt}{0.4pt}}}\ULon}

\newcommand{\beq}{\begin{equation}}
\newcommand{\eeq}{\end{equation}}
\newcommand{\beqar}{\begin{eqnarray}}
\newcommand{\eeqar}{\end{eqnarray}}
\newcommand{\bitm}{\begin{itemize}}
\newcommand{\eitm}{\end{itemize}}

\begin{document}


\title{{\bf \boldmath Updated measurements of the M1 transition $\psp\to\gamma\etacp$ with $\etacp\to K\bar{K}\pi$ }}

\author{M.~Ablikim$^{1}$, M.~N.~Achasov$^{5,b}$, P.~Adlarson$^{75}$, X.~C.~Ai$^{81}$, R.~Aliberti$^{36}$, A.~Amoroso$^{74A,74C}$, M.~R.~An$^{40}$, Q.~An$^{71,58}$, Y.~Bai$^{57}$, O.~Bakina$^{37}$
% ...
\\
\vspace{0.2cm}
(BESIII Collaboration)\\
\vspace{0.2cm} {\it
$^{1}$ Institute of High Energy Physics, Beijing 100049, People's Republic of China\\
% ...
$^{5}$ Budker Institute of Nuclear Physics SB RAS (BINP), Novosibirsk 630090, Russia\\
% ...
$^{36}$ Johannes Gutenberg University of Mainz, Johann-Joachim-Becher-Weg 45, D-55099 Mainz, Germany\\
$^{37}$ Joint Institute for Nuclear Research, 141980 Dubna, Moscow region, Russia\\
% ...
$^{40}$ Liaoning Normal University, Dalian 116029, People's Republic of China\\
% ...
$^{57}$ Southeast University, Nanjing 211100, People's Republic of China\\
$^{58}$ State Key Laboratory of Particle Detection and Electronics, Beijing 100049, Hefei 230026, People's Republic of China\\
% ...
$^{71}$ University of Science and Technology of China, Hefei 230026, People's Republic of China\\
% ...
$^{74}$ University of Turin and INFN, (A)University of Turin, I-10125, Turin, Italy; (B)University of Eastern Piedmont, I-15121, Alessandria, Italy; (C)INFN, I-10125, Turin, Italy\\
$^{75}$ Uppsala University, Box 516, SE-75120 Uppsala, Sweden\\
% ...
$^{81}$ Zhengzhou University, Zhengzhou 450001, People's Republic of China\\
\vspace{0.2cm}
% ...
$^{b}$ Also at the Novosibirsk State University, Novosibirsk, 630090, Russia\\
% ...
}}


\vspace{0.4cm}
\date{\today}
\begin{abstract}

Based on a data sample of $(27.08 \pm 0.14 )  \times 10^8~\psp$ events collected with the BESIII detector at the BEPCII collider, 
the M1 transition $\ptge$ with $\etacp\to K\bar{K}\pi$ is studied, where $K\bar{K}\pi$ is $\kkp$ or $\kskp$.
The mass and width of the $\etacp$ are measured to be $(3637.8 \pm 0.8 (\rm {stat}) \pm 0.2 (\rm {syst}))~\mevcs$ and $(10.5 \pm 1.7 (\rm {stat}) \pm 3.5 (\rm {syst}))~\mev$, respectively.
The product branching fraction $\mathcal{B}\left(\psi(3686) \rightarrow \gamma \eta_{c}(2 S)\right) \times \mathcal{B}(\eta_{c}(2 S) \rightarrow K \bar{K} \pi)$ is determined to be $(0.97 \pm 0.06  (\rm {stat}) \pm 0.09 (\rm {syst})) \times 10^{-5}$. Using $\BR(\etacp\to K\bar{K}\pi)=(1.86^{+0.68}_{-0.49})\%$, we obtain the branching fraction of the radiative transition to be $\BR(\ptge) = (5.2 \pm 0.3  (\rm {stat}) \pm 0.5 (\rm {syst}) ^{+1.9}_{-1.4}  (extr)) 
\times 10^{-4}$, where the third uncertainty is due to the quoted $\BR(\etacp \to K\bar{K}\pi)$.
\end{abstract}

\pacs{Valid PACS appear here}
\maketitle

\section{Introduction}
\label{sec:introduction}

Charmonia, comprising bound states of a charmed quark and its antiquark~($c\bar{c}$),  provide a unique opportunity to study the strong interactions in the $\tau$-charm energy regime, where non-perturbative contributions are comparable to perturbative ones. In the low energy region, the large coupling of the strong interaction ($\alpha_s$) makes first-principle calculations extremely difficult.  Thus, various theoretical models have been developed~\cite{Barnes:2005pb,Li:2007xr,Peng:2012tr,Brambilla:2010cs}, and good agreement has been achieved for the mass spectrum of charmonia below the open-charm threshold. However, long-standing puzzles in charmonium physics still exist, such as the $\rho-\pi$ puzzle~\cite{rho-pi1,rho-pi2,whp} and substantial non-$D\bar{D}$ decays of the $\psi(3770)$~\cite{DD:zq,BES:zyz}. Further studies of the charmonium system are desired, both theoretical and experimental, to resolve these puzzles and consequently better understand the strong interaction. 

The charmonium spin-singlet state $\eta_{c}(2S)$ was first observed by the Belle collaboration in the $B$ meson decay, $B^{\pm}\rightarrow K^{\pm} \eta_{c}(2S)$, with $\eta_{c}(2S)\rightarrow \kskp$ in 2002~\cite{bell:etac}. However, knowledge about this particle is still very limited. For example, the sum of its branching fractions (BFs) measured experimentally is only about 3\%~\cite{PDG}.
Since the $\eta_{c}(2S)$ is a pseudo-scalar particle with quantum numbers $J^{PC} = 0^{-+}$, it cannot be produced directly via electron-positron annihilation.
The production of the $\eta_{c}(2S)$ through $\psi(3686)$ radiative transition requires a charmed-quark spin-flip and, thus, proceeds via a magnetic transition. The partial width $\Gamma(\ptge)$~\cite{Barnes:2005pb,Li:2007xr,Peng:2012tr} and the BF $\BR(\ptge)$~\cite{oth_the} have been calculated in different theoretical frameworks, as shown in Table~\ref{Tab:Width_theoretical}. These include the non-relativistic potential model (NR model)~\cite{Barnes:2005pb,Li:2007xr}, the Godfrey-Isgur relativized 
potential model (GI model)~\cite{Barnes:2005pb,Li:2007xr}, the light-front quark model~\cite{Peng:2012tr}, and effective field theory~\cite{oth_the}, among others. Significant discrepancies exist among the predictions. 

\begin{table*}[htbp]
\centering
\footnotesize
\setlength{\tabcolsep}{6pt}
\renewcommand{\arraystretch}{1.2}
\caption{A comparison of theoretical calculations for 
the $\eta_c(2S)$ mass, the BF $\BR(\ptge)$, and the partial width $\Gamma(\ptge)$.}
\begin{tabular}{c c c c}
\hline\hline 
     &Mass~($\mevcs$)     &$\BR(\ptge)~(\times10^{-4})$  &$\Gamma(\ptge)~(\kev)$\\
\hline
NR model~\cite{Barnes:2005pb} &3630  & $7.14\pm 0.19$  & 0.21\\
GI model~\cite{Barnes:2005pb} &3623  & $5.80\pm0.16$   & 0.17\\
Meson loop correction~\cite{Li:2007xr}          & N/A  & $2.72 \pm 1.00$  &$0.08\pm0.03$\\
Light-front quark model~\cite{Peng:2012tr}       &3637  & 3.9             &0.11 \\
Effective field theory~\cite{oth_the}            &N/A  & $0.6-36.0$       & N/A \\
\hline \hline
\end{tabular}
\label{Tab:Width_theoretical}
\end{table*}
 
On the experimental side, the first observation of the M1 transition $\ptge$ with $\etacp$ decaying 
into $K\bar{K}\pi$ was reported by BESIII~\cite{BES:2012uhz}. With a data sample of 
$106 \times 10^{6}~\psi(3686)$ events, BESIII measured the BF $\BR(\ptge) = (6.8\pm 1.1 \pm 4.5) \times 10^{-4}$, 
where, here and hereafter, the first uncertainty is statistical and the second is systematic. 
Reference~\cite{whp} extracted $\BR(\ptge) = (7^{+3.4}_{-2.5})\times 10^{-4}$ by a global fit to many experimental measurements and is in good agreement with the world average value of $(7 \pm 5)\times 10^{-4}$~\cite{PDG}. It is difficult to validate the theoretical calculations with the current measurements, since the relative uncertainties are as large as 70\%. 

The mass of the $\etacp$ was predicted to be $3.630~\gevcs$, $3.623~\gevcs$, and $3.637~\gevcs$ according to the NR model~\cite{Barnes:2005pb}, 
the GI model~\cite{Barnes:2005pb}, and in Ref.~\cite{Peng:2012tr}, respectively.  Recently, the LHCb collaboration reported the mass of the $\etacp$ to be $m_{\etacp} = (3637.90 \pm 0.54 \pm 1.40 )~\mevcs$ and its width to be $\Gamma_{\etacp} = (10.77 \pm 1.62 \pm 1.08)~\mev$, via $B \rightarrow K \etacp, \etacp \rightarrow K_{S}^{0}K\pi$~\cite{CHLb}. The experimental averages of its resonant parameters are 
$m_{\etacp} = (3637.5 \pm 1.1)~\mevcs$ and $\Gamma_{\etacp} = (11.3^{+3.2}_{-2.9})~\mev$~\cite{PDG}. 

The BESIII experiment has collected $(27.08 \pm 0.14 ) \times 10^8~\psp$ events~\cite{Bes:data2021},  which is 25.5 times larger than the dataset used in 
the previous measurement~\cite{BES:2012uhz}. Using this enlarged data sample, we present a comprehensive analysis of the M1 transition 
$\ptge$ with $\etacp$ decaying into $K\bar{K}\pi$ final states. The resonant parameters of the $\etacp$ and the 
BF of $\ptge$ are reported with improved precision.

\section{Detector and Monte Carlo simulation}
\label{detector}

The BESIII detector~\cite{Ablikim:2009aa} records symmetric $e^+e^-$ collisions 
provided by the BEPCII storage ring~\cite{BESIII:2020nme}
in the center-of-mass energy range from 2.0 to 4.95~GeV,
with a peak luminosity of $1 \times 10^{33}\;\text{cm}^{-2}\text{s}^{-1}$ 
achieved at $\sqrt{s} = 3.77\;\text{GeV}$.  BESIII has collected large data samples in this energy region~\cite{BESIII:2020nme}. The cylindrical core of the BESIII detector covers 93\% of the full solid angle and consists of 
a helium-based multilayer drift chamber~(MDC), a plastic scintillator time-of-flight system~(TOF), and a CsI(Tl) 
electromagnetic calorimeter~(EMC), which are all enclosed in a superconducting solenoid magnet providing a 1.0~T 
magnetic field. The solenoid is supported by an octagonal flux-return yoke with resistive plate counter muon 
identification modules interleaved with steel. 
The charged-particle momentum resolution at $1~\text{GeV}/c$ is $0.5\%$, and the $dE/dx$ resolution is 6\% for 
electrons from Bhabha scattering. The EMC measures photon energies with a resolution of 2.5\% (5\%) at $1~\gev$ in 
the barrel (end cap) region. The time resolution in the TOF barrel region is $68~\ps$, while that in the end cap region 
is $110~\ps$. The end cap TOF system was upgraded in 2015 using multi-gap resistive plate chamber technology, providing a 
time resolution of $60~\ps$~\cite{etof}. 

Simulated data samples produced with a {\sc
geant4}-based~\cite{geant4} Monte Carlo (MC) package, which
includes the geometric description of the BESIII detector~\cite{BESIII:detector_descrip} and the detector response, are used to determine the detection efficiencies and to estimate backgrounds. The simulation models the beam energy spread and initial state radiation (ISR) in 
the $\EE$ annihilation with the {\sc kkmc} generator~\cite{ref:kkmc}. The inclusive MC sample includes the production of 
the $\psp$ resonance, the production of $\jpsi$ via ISR, and the continuum processes incorporated in {\sc kkmc}. All the known particle decays are modeled with {\sc evtgen}~\cite{ref:evtgen} using the BFs either taken from 
the Particle Data Group~\cite{PDG}, when available,  or otherwise estimated with {\sc lundcharm}~\cite{ref:lundcharm}. 
The final state radiation~(FSR) from charged final state particles is incorporated using the {\sc photos} package~\cite{photos}. We 
generate the $\ptge$ decay using helicity amplitudes~\cite{BES:suxian} and the $\etacp \to K\bar{K}\pi$ decays with a phase 
space~(PHSP) model.

\section{Event selection}
\label{sec:selection}
In selecting signal events for the process $\psp\to\gamma \kkpII$, we require each candidate event to contain four~(two) charged tracks 
with zero net charge, and at least one~(three) photon(s) for the $\gamma \kskp~(\gamma \kkp)$ mode. Here, $\kkpII$ represents only $\kkp$ and $\kskp$ in Secs.~\ref{sec:selection},~\ref{sec:bg}, and~\ref{sec:br} for simplicity except explicitly stated. We form $\ks$ ($\piz$) candidates using pairs of $\pp$~($\GG$).

Photon candidates are identified using showers in the EMC. The deposited energy of the shower must be more than $25~\mev$ 
in the barrel region ($|\cos \theta|< 0.80$) or more than $50~\mev$ in the end cap region ($0.86 <|\cos \theta|< 0.92$).
To exclude showers that originate from charged tracks, the angle subtended by the EMC shower and the position of the 
closest charged track at the EMC must be greater than 10 degrees as measured from the interaction point (IP). To suppress the 
electronic noise and the showers unrelated to the event, the difference between the EMC time and the event start time 
is required to be within $[0,~700]~\ns$.

A charged track is reconstructed from the hits in MDC. We require each charged track not originating from $\ks$ to 
satisfy $|\cos \theta|<0.93$, and the distance of the closest approach to the IP must be within $10~\cm$ along the $z$-axis, and less than
$1~\cm$ in the transverse plane. 
Particle identification~(PID) for charged tracks combines measurements of the energy deposited in the MDC~(d$E$/d$x$) and the flight time in the TOF to form likelihoods $\mathcal{L}(h)~(h=p,K,\pi)$ for each hadron $h$ hypothesis. Tracks are identified as $K^{\pm}$ or $\pi^{\pm}$ by comparing the likelihoods for the kaon and pion hypotheses, $\mathcal{L}(K)>\mathcal{L}(\pi)$ and $\mathcal{L}(\pi)>\mathcal{L}(K)$, respectively.

Each $\ks$ candidate is reconstructed from two oppositely charged tracks which are assigned as $\pp$ without imposing further PID criteria. 
They are constrained to
originate from a common vertex and are required to have an invariant mass
within $|M_{\pi^{+}\pi^{-}} - m_{K_{S}^{0}}|<$ 7~MeV$/c^{2}$, where
$m_{K_{S}^{0}}$ is the $K^0_{S}$ nominal mass~\cite{PDG}. Here the $\ks$ signal has a mass resolution of $3.5~\mevcs$. The
decay length of the $K^0_S$ candidate is required to be greater than
twice the vertex resolution away from the IP.
  
For $\gamma\kskp$ candidate events, we 
perform a four-constraint ($4C$) kinematic fit to all the final state particles with the constraints provided by 
four-momentum conservation, where the $\ks$ candidate information is from the secondary-vertex fit. We discriminate the $\ks K^+ \pi^-$ and $\ks K^- \pi^+$ charge-conjugate combinations and 
select the best photon candidate by minimizing $\chi^2_{com} = \chi^2_{4C} + \chi^2_{\rm PID}(K) + 
\chi^2_{\rm PID}(\pi)$, where $\chi^2_{4C}$ is from the $4C$ kinematic fit. For the $\gamma\kk\piz$ mode, we 
use a five-constraint $(5C)$ kinematic fit with an additional constraint on the $\piz$ nominal mass and select 
the combination with the minimum $\chi^2_{5C}$. To suppress backgrounds from $\psp\to \kkpII$ and $\psp \to \gamma \gamma \kkpII$, we require the $\chi^2$ of 
the kinematic fit of the $\gamma \kkpII$ hypothesis to be less than that from both the $\kkpII$ and $\GG \kkpII$ hypotheses.

\section{Backgrounds analysis}
\label{sec:bg}
For the $\kskp$ mode, there are significant background contributions from $\psp\to\pp\jpsi$ with $\jpsi$ decaying into $\EE$ and $\MM$, $\psp \to \eta \jpsi$ with $\eta$ decaying into $\ppp$, and from $\psp \to \gamma \ks\ks$ events with the $K^{\pm}$ being misidentified as $\pi^{\pm}$. The other backgrounds are from $\psp\to \piz \jpsi$ or $\psp\to \piz \piz \jpsi$ with $\jpsi$ decaying into $\EE$, $\MM$, or $\kk$, $\psp \to \eta \jpsi$ with $\eta$ decaying into $\GG$, $\psp \to \omega \kk$ with $\omega \to \gamma 
\piz\to 3\gamma$ for the $\kkp$ mode.
To suppress these backgrounds, we require the recoil mass of 
all the $\pp$ pairs to be less than $3.05~\gevcs$ related $\jpsi$ backgrounds, and the two charged tracks other than those from the $\ks$ candidate must have an invariant mass differing from $m_{\ks}$ by at least $10~\mevcs$ for $\gamma \ks\ks$ backgrounds in the $\kskp$ mode. The invariant mass of the two charged tracks with 
$\mu^\pm$ hypothesis to be less than $2.90~\gevcs$ for the $\kkp$ mode. We suppress 
the background from $\psp \to \omega \kk$, in which $\omega \to \gamma \pi^0 \to 3\gamma$, using an $\omega$ veto in the $\kkp$ mode, i.e., the invariant mass ($M_{3\gamma}$) 
of the $3\gamma$ must satisfy $M_{3\gamma} < 0.74~\gevcs$ or $M_{3\gamma} > 0.82~\gevcs$. 

After imposing the above requirements, the analysis of the inclusive MC sample for $\psp$ decays with TopoAna~\cite{BES:topo} 
indicates that the remaining dominant background sources are: (1) $\psp \to \kkpII$ events with a fake photon ($\gfake$) or a photon 
from FSR in the final state; (2) events with an extra photon, primarily from $\psp \to \piz \kkpII$ with $\piz\to\GG$; 
and (3) events from continuum process $e^{+}e^{-} \rightarrow \gisr(\gfsr) \kkpII$ with the photon from ISR or FSR.  

\subsection{$\psp\to \kkpII$ with a $\gfake$ or FSR}

\begin{figure}[htb]
\centering
\subfigure{\includegraphics[width=8cm]{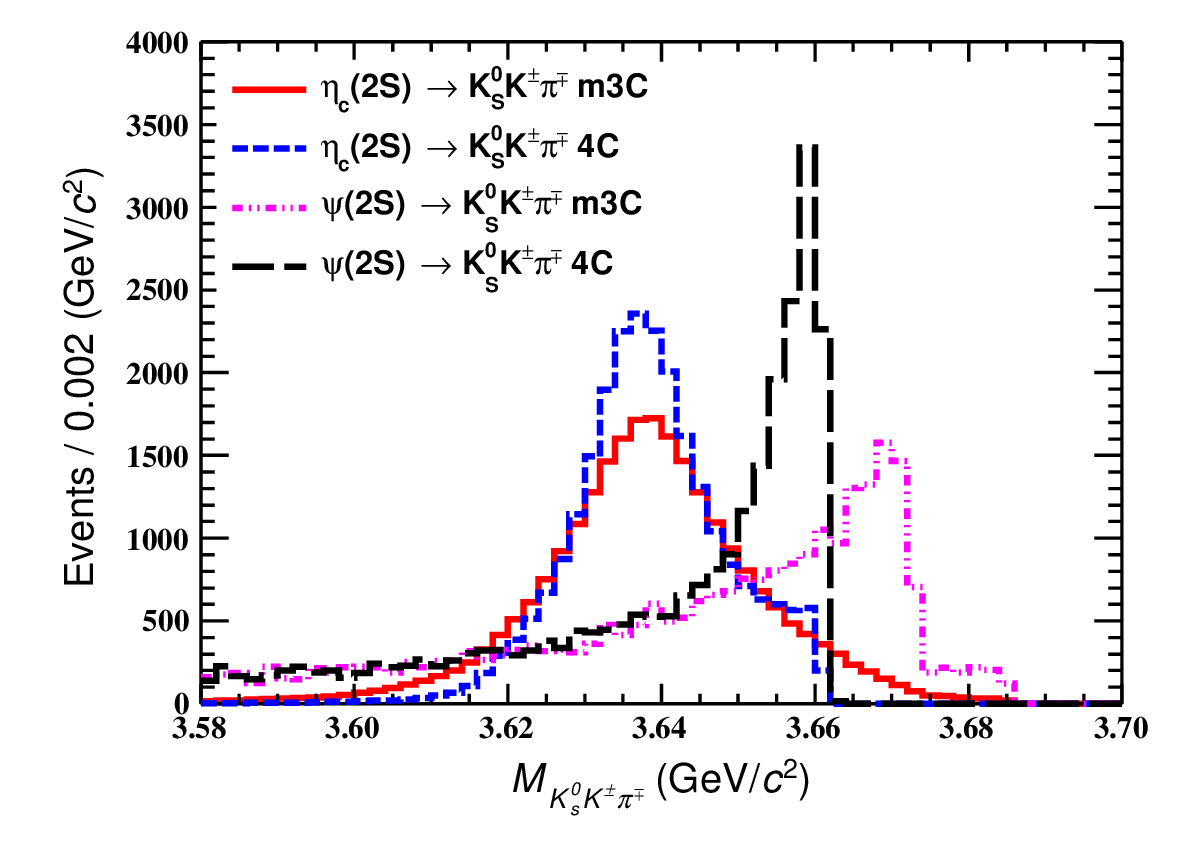}}
 \put(-50,140){\bf ~(a)}\\
\subfigure{\includegraphics[width=8cm]{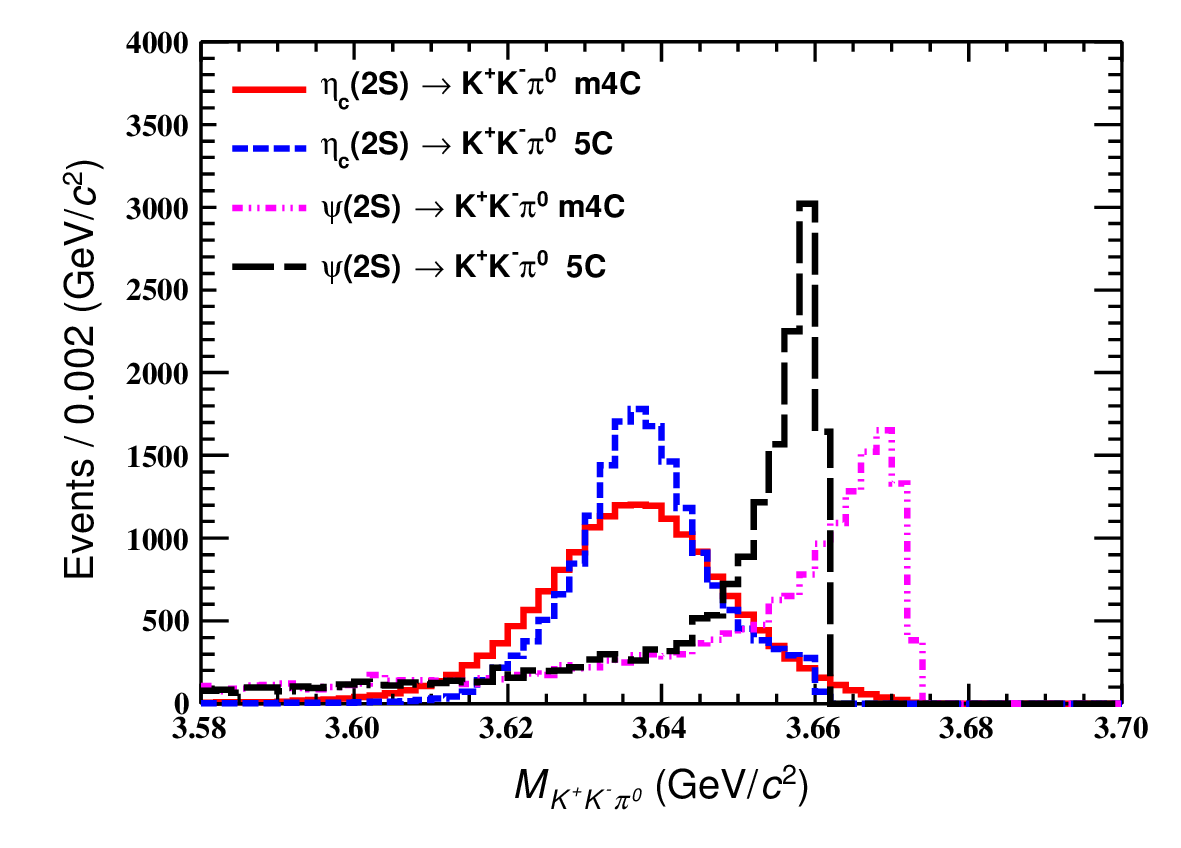}}
 \put(-50,140){\bf ~(b)}\\
\caption{A comparison between the $4C$ ($5C$) and the modified $3C$ ($4C$) kinematic fits of the MC samples. 
(a) and (b) are the invariant mass distributions of $\kskp$ and $\kkp$ from $\etacp$ or $\psp$ decays, respectively. }
\label{3c4c}
\end{figure}

We show the distributions of the invariant mass of $\kkpII$ ($M_{\kskp}$ and $M_{\kk\piz}$) after the kinematic fits in Fig.~\ref{3c4c}, where 
the backgrounds from $\psp \to \kkpII$ with a $\gfake$ incorporated into the kinematic fit, appear as peaks close to the expected $\etacp$ mass, with a sharp cutoff due to the 25-MeV photon-energy threshold. Since the $\gfake$ adds no information to the fit, its inclusion distorts the mass measurement.
A modified kinematic fit is therefore imposed by allowing the energy of the photon to float, i.e., using a 
$3C$ kinematic fit for $\gamma \kskp$ candidates and a $4C$ kinematic fit for $\gamma \kkp$ candidates. We find the energy of the $\gamma$ 
from the kinematic fit tends to be zero if it is a fake photon, which is useful for better separating such kinds of backgrounds from the signal.  We require the $\chi^2$ in the modified kinematic fits to satisfy $\chi^2_{m3C} < 20$ and $\chi^2_{m4C} < 15$ for the $\gamma \kskp$ mode and the $\gamma \kkp$ mode, 
respectively. We use the invariant mass distributions from the modified fits for further study.

The above requirement cannot remove backgrounds from $\psp \to \kkpII$ with a $\gfsr$, which could potentially contaminate the 
signal channel with a long tail in the distribution of $M_{K\bar{K}\pi}$. The contribution depends on the FSR ratio 
$R_{\rm FSR} \equiv N_{\rm FSR} / N_{\rm noFSR}$ ignoring any dependence on the momentum and angle of the charged tracks, where $N_{\rm FSR}$ and 
$N_{\rm nonFSR}$ are the numbers of events with and without $\gfsr$, respectively. Control samples of $\psp \to \gamma 
\chico \to \gamma \kskp (\gfsr)$ and $\psp \to \gamma \chicz \to \gamma \kk\kk (\gfsr)$ are selected to study the difference 
of $R_{\rm FSR}$ between data ($R_{\rm FSR}^{\rm Data}$) and MC sample ($R_{\rm FSR}^{\rm MC}$) in the $\gamma K_{S}^{0} K^{\pm} \pi^{\mp}$ and $\gamma K^{+} K^{-} \pi^{0}$ channels, respectively. We use the same method as described in Ref.~\cite{BES:2012uhz} to determine $f_{\rm FSR} 
\equiv R_{\rm FSR}^{\rm Data}/R_{\rm FSR}^{\rm MC} = (1.53 \pm 0.05)$ and $(1.34\pm0.12)$ from two different control samples.
We tune the ratio of $\psp \to \kkpII$ and $\psp \to \gfsr \kkpII$ in the MC simulation according to $f_{\rm FSR}$ to determine the $M_{\kkpII}$ line-shape of $\psp \to (\gfsr) \kkpII$, which will be used to describe the corresponding background contribution in the fit.

\subsection{$\psp \to \piz  \kkpII$}
To account for $\psp \to \piz  \kkpII$ backgrounds,
we use the same data-driven method as applied in Ref.~\cite{BES:2012uhz}. We first select $\psp \to \piz  \kkpII$ events in data and then estimate their rate of contamination in the $M_{\kkpII}$ spectrum by exposing the selected events to the $\gamma \kkpII$ selection criteria and correcting with efficiencies determined from MC simulation.

This background contributes a smooth component around the $\chi_{c1,2}$ mass region with a small tail in the $\etacp$ 
signal region. We fit these backgrounds with a Novosibirsk function~\cite{Novosibirsk}, and 
fix its shape and size in the $M_{\kkpII}$ spectra fits.

\subsection{ISR and FSR events from $\EE$ annihilation}

We use data collected at $\sqrt{s} = 3.65~\gev$ with a luminosity of $401.00\pm4.01$ pb$^{-1}$~\cite{Bes:data2021} to estimate the backgrounds from the continuum processes $\EE \to \gfsr 
K\bar{K}\pi $ and $\EE \to \gisr K\bar{K}\pi$. After event selection, we shift the mass of $K\bar{K}\pi~(M_{\rm shifted})$ to $3.686~\gev$ by the rule:
\beq
M_{\rm shifted}=a \times (M_{K\bar{K}\pi}-m_{0})+m_{0} ,
\eeq
where $a=(3.686-m_{0})/(3.65-m_{0})$ with $m_{0}$ being the mass threshold for generating $K\bar{K}\pi$. We then normalized the number of continuum backgrounds according to the cross sections and the 
integrated luminosities of the data samples at $\sqrt{s} = 3.65~\gev$ and at the $\psp$ peak~\cite{Bes:data2021}. We determine the number
of continuum events after the selection in the $\psp$ data sample to be 516 (360) in the $\gamma \kskp$ ($\gamma \kkp$) final state. 
 
\begin{figure*}[htb]
\centering
\includegraphics[angle=0,width=0.9\textwidth]{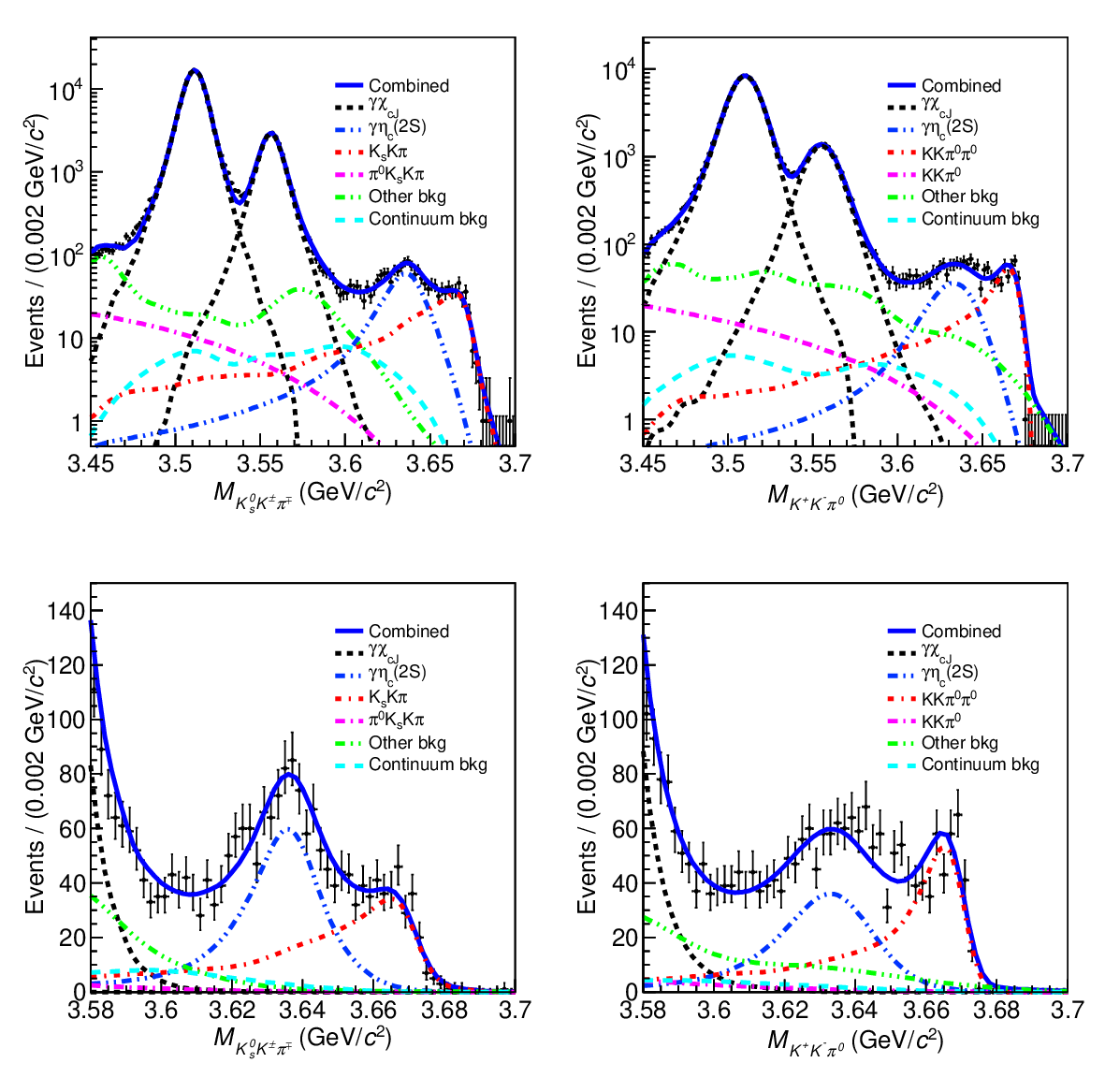}
\put(-415,415){\bf ~(a)}
\put(-185,415){\bf ~(b)}
\put(-415,185){\bf ~(c)}
\put(-185,185){\bf ~(d)}
\caption{ The distributions of the invariant mass spectra of $K\bar{K}\pi$. The fit results in the whole fit range (a,b) and in the range only containing $\etacp$ signal (c,d). (a) and (c) are for the $\kskp$ mode, (b) and (d) are for the $\kkp$ mode. Dots with error bars are data, blue solid curves are the best results from the 
simultaneous fits. The blue dashed lines are the $\etacp$ signals, and the other dashed lines are backgrounds, among 
which the pink, red, black, green and cyan ones are from the $\psp\to \piz+K\bar{K}\pi$ events, the $\psp\to K\bar{K}\pi$ events, 
the $\psp\to\gamma\chi_{c1,2}$ events, other contributions, and continuum contribution, respectively. For plots (a) and (c), the data statistics is summed over the two charge conjugated modes. }
\label{Fig:sim_fit}
\end{figure*}
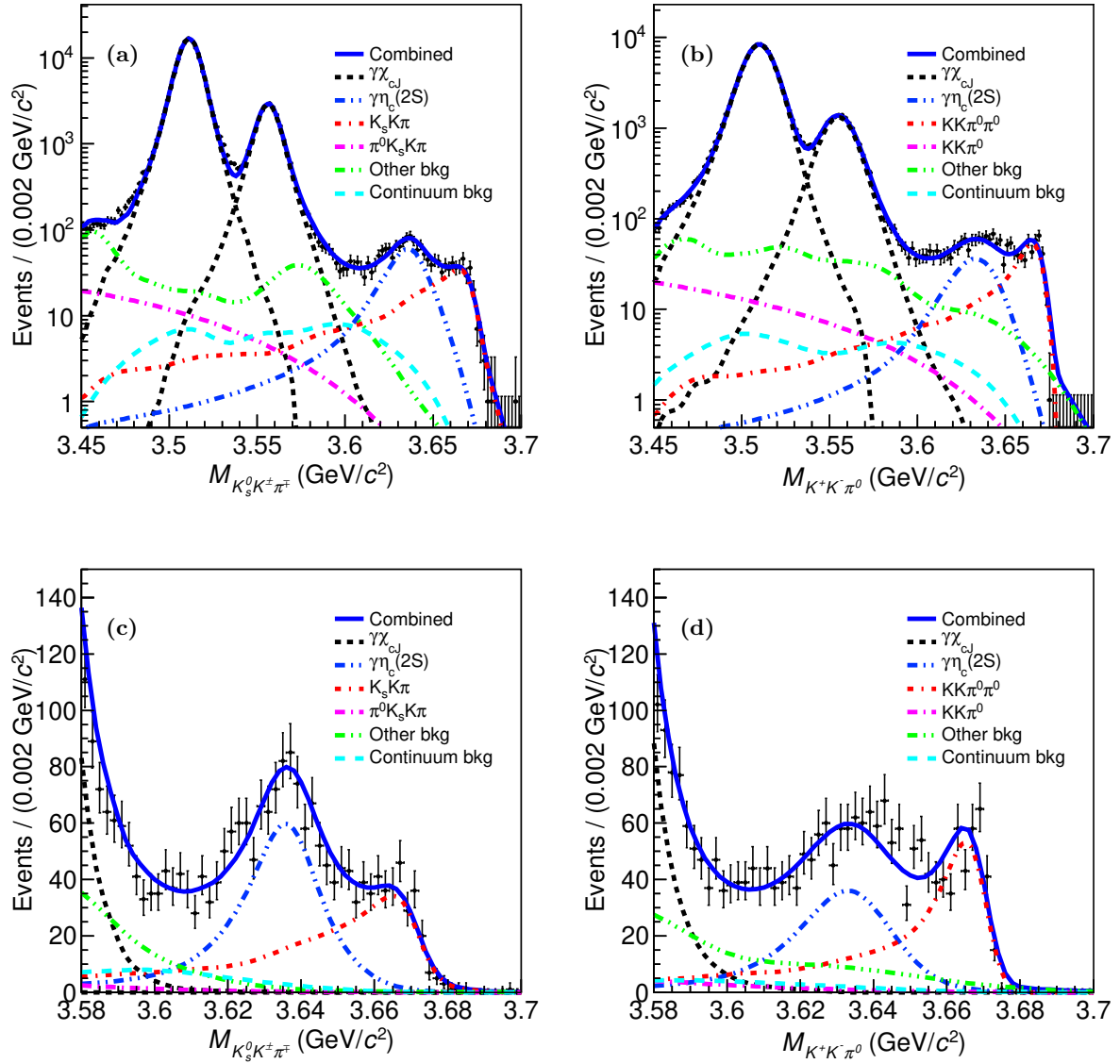
 
\section{Branching fraction determination} 
\label{sec:br}
We perform a simultaneous fit to the mass spectra of $K\bar{K}\pi$ in the range of $(3.45-3.70)~\gevcs$ to extract the yield of 
signal events and the mass and the width of the $\etacp$, as shown in Fig.~\ref{Fig:sim_fit}. There are $\chico$ and $\chict$ 
signals besides the $\etacp$ signals.
We use the following line-shape of the $\etacp$ produced by an M1 transition:
\beq
(E_\gamma^3 \times BW(M_{K\bar{K}\pi}) \times f_{d}\left(E_\gamma\right) \times \eff(M_{K\bar{K}\pi}) ) \otimes \operatorname{DGaus},
\eeq
where $E_\gamma = (m_{\psp}^2-M^2_{K\bar{K}\pi})/(2 m_{\psp})$ is the energy of the transition photon in the rest frame of the $\psp$, $BW$ is a Breit-Wigner function with floating width and mean,  
$f_d (E_\gamma) = E_0^2/(E_\gamma E_0 + 
(E_\gamma - E_0)^2)$ is a damping function proposed by the KEDR experiment~\cite{KEDR} to suppress the diverging tail raised by the term of 
$E^3_\gamma$ with $E_0 = (m^2_{\psp} - m^2_{\etacp})/(2 m_{\psp})$ denoting the mean energy of the transition photon, 
$\eff(M_{K\bar{K}\pi})$ is the efficiency as a fitted polynomial function of $M_{K\bar{K}\pi}$ from the MC simulation,
 $\rm DGaus$ is a double Gaussian function, i.e. $f\cdot Gaus(m_1, \sigma_1) + (1-f)\cdot Gaus(m_2,\sigma_2)$, that is utilized to describe the resolution. In fact, there are two types of resolutions that need to be taken into account. The first type is based on the detector resolution, which is represented by a double Gaussian function. The parameters of this function are directly determined through MC simulation. The second type of resolution is based on the discrepancy between the data and MC simulation, and is represented by a single Gaussian function. The parameters of this function are obtained through a linear extrapolation from those of $\chi_{c1/2}$. To consider both types of resolutions, a new double Gaussian function, $\rm DGaus$,  is constructed by convoluting the original double Gaussian function with the single Gaussian function. All parameters of $\rm DGaus$ are then determined and fixed during the fitting process.

The line-shapes of the $\chi_{c1,2}$ are described with the simulated MC shapes convoluted 
with a single Gaussian function for the difference in resolution of $M_{K\bar{K}\pi}$ between data and MC simulation, and their parameters are floated in the fit.

We use a Novosibirsk function~\cite{Novosibirsk} to parameterize the backgrounds of $\psp\to \pi^{0}K\bar{K}\pi$ events from data, with the shape of the function 
and the number of background events fixed. For the background of $\psp \to K\bar{K}\pi$ with and without $\gfsr$, the shape 
is extracted from MC simulation, with the FSR and non-FSR components reweighted by corrected $f_{\rm FSR}$.
We add the continuum backgrounds contribution with shifted mass spectrum shape, and the numbers of continuum backgrounds are fixed.
We describe all other smooth backgrounds with the shape fixed to that from the inclusive MC sample while the number of events are floated.  

The yields of the $\etacp$ signal are $692\pm 42$ in the $\kkp$ mode and $963 \pm 58$ in the $\kskp$ mode. 
From the signal MC simulations, we obtain the efficiency  $\eff_{\kkp} = 16.02\%$ in the $\kkp$ mode and $\eff_{\kskp} 
= 15.93\%$ in the $\kskp$ mode, respectively. In the MC simulation, the mass and width of the $\eta_{c}(2S)$ are quoted from the world averaged values~\cite{PDG}. Therefore, we determine the products of the BFs to be $\BR(\ptge) 
\times \BR(\eta_c(2S) \to K^0_S K^+ \pi^- + c.c.) = (3.23 \pm 0.20) \times 10^{-6}$ and $\BR(\ptge) \times \BR(\etkkp)= (1.61 \pm 0.10) 
\times 10^{-6}$. The ratio of the BFs agrees well with the isospin symmetry expectation of 2:1 between $\kskp$ and $\kkp$. We determine the mass and width of the $\etacp$ to be $M_{\etacp} = (3637.8 \pm 0.8 )~\mevcs$ and 
$\Gamma_{\etacp} = (10.5\pm 1.7)~\mev$. The uncertainties are statistical only.

Considering isospin conservation, the product BF for $\ptge$ with $\etacp\to K\bar{K}\pi$ can be obtained 
by doubling the sum of the $\kskp$ and $\kkp$ BFs to obtain $\BR(\ptge)\times \BR(\etacp \to K\bar{K}\pi) = 
(0.97 \pm 0.06) \times 10^{-5}$, where the uncertainty takes into account the correlation between the two measured 
BFs from the simultaneous fit. Here $K\bar{K}\pi$ indicates the sum of seven channels $K^+ K^- \pi^0$, $ K_SK_S\pi^0$, $K_LK_L\pi^0$, $K_SK^+\pi^-$, $K_SK^-\pi^+$, $K_LK^+\pi^-$, $K_LK^-\pi^+$. If only consider the isospin symmetry, the ratios of the BFs between them should be $2:1:1:2:2:2:2$.

\section{systematic uncertainties}
\label{sec_sys}

There are multiple sources of systematic uncertainties in determining the resonant parameters of the $\etacp$ and the product of branching 
fractions $\BR(\ptge)\times \BR(\etacp \to K\bar{K}\pi)$.

\begin{table*}[htb]
\setlength{\abovecaptionskip}{5pt} 
\setlength{\belowcaptionskip}{4pt}
\centering
\setlength{\tabcolsep}{7pt}
\renewcommand{\arraystretch}{1.2}
\caption{The systematic uncertainty of the $\etacp$ mass and width, and the systematic uncertainty of the product 
BF $\BR(\ptge) \times \BR(\etacp \to K \bar{K} \pi)$, represented by $\BR\BR$ in this Table.}
\begin{tabular}{ccccc}
\hline \hline  
& Source & Mass ($\mevcs$) & Width ($\mev$) & $\BR\BR$ (\%) \\
\hline \multirow{7}{*}{$K\bar{K}\pi$ event selection } & Tracking & $-$ &$-$ & 2.0\\
& PID                      & $-$          &$-$     & 2.0\\
& Photon reconstruction    & $-$          &$-$     &2.7 \\
& Kinematic fit            & $-$          &$-$     &4.3 \\
& $K_{S}^{0}$ reconstruction& $-$         &$-$     &0.8 \\
& Resonance parameters of $\etacp$  & $-$   &$-$     &1.8 \\
& Intermediate resonance  & $-$   &$-$     &2.9 \\
\hline \multirow{5}{*}{ Simultaneous Fit } & Resolution function                  & 0.1             & 0.3     & 0.6  \\
& Efficiency curve        & 0.0           & 0.0     & 0.1 \\
& Background line-shape   & 0.0           & 2.9     & 3.2 \\
&Damping function        & 0.1           & 1.9     & 4.5 \\
&Isospin constraint       &0.0            & 0.1     &$-$   \\
\hline $\psi(3686)$ data sample & &$-$    &$-$       &0.5 \\
\hline Number of continuum events &&0.0   &0.2     &1.0 \\
\hline Shape of continuum   &&0.1         &0.5     &2.0 \\
\hline & Combined           &0.2          &3.5     &9.1 \\
\hline \hline 
\end{tabular}
\label{Tab:fit_sysunc}
\end{table*}

Systematic uncertainties associated with event selection, such as tracking, PID, photon reconstruction, $\ks$ reconstruction, and kinematic fit are all estimated using control samples. To account for each source's uncertainty, we vary the efficiency accordingly and calculate the difference in the final BF. This difference represents the systematic uncertainty attributed to tracking, PID, and photon reconstruction, respectively. From the study with two control samples, $\jpsi \to p \bar{p} \pp$ and $\EE \to \pp\kk$, the uncertainties of tracking and PID are $1\%$, respectively, for each pion and kaon track~\cite{track_sys1,track_sys2}. 
According to a study with the control sample $\jpsi\to \rho\pi$, the uncertainty due to photon detection efficiency is 
1\% per photon and is additive~\cite{BESIII:gamma_sys}.
The systematic uncertainty from the $\ks$ reconstruction is studied using two control samples, $\jpsi\to K^{*\pm}(892) K^{\mp}$ 
with $K^{*\pm}(892) \to \ks \pi^{\pm}$ and $\jpsi \to \phi \ks K^{\pm} \pi^{\mp}$~\cite{BESIII:Ks_sys}. We estimate the  
systematic uncertainty from $\ks$ reconstruction to be 1.2\% for the $\kskp$ mode.
To study the uncertainty associated with the kinematic fit, we correct the track helix parameters in the MC 
simulation~\cite{helix}. The efficiency difference 2.8\% and 5.1\% before and after the correction is taken as 
the systematic uncertainty related to the kinematic fit for the $\kkp$ mode and $\kskp$ mode, respectively. The resonance parameters of $\eta_{c}(2S)$ and the intermediate resonances also impact the event selection efficiency. We estimate the systematic uncertainty associated with the resonance parameters of $\eta_{c}(2S)$ by varying its mass and width by one standard deviation relative to the world average values~\cite{PDG}. Additionally, we consider the influence of intermediate resonances by generating new mixed MC samples with additional decay channels, namely $\eta_{c}(2S)\to K_{0}^{*}(1430)\bar{K}$ and $\eta_{c}(2S)\to K_{2}^{*}(1430)\bar{K}$, using fractions inspired by the LHCb study~\cite{CHLb}. The difference in efficiency between the new mixed MC samples and the nominal MC samples is taken as the corresponding systematic uncertainty, which is estimated to be 2.9\%.

The uncertainties related to the simultaneous fit include those due to the signal resolution function, the efficiency curve, 
the line-shapes of the backgrounds, and the damping function. We estimate the systematic uncertainty due to the resolution function by changing the parameters of the mass and the 
mass resolution in the single Gaussian function and the convolved double Gaussian function by $1\sigma$. The largest 
difference is 0.6\%, and we take it as the uncertainty due to the resolution function.
The uncertainty from the efficiency curve is estimated by comparing the difference in results with and without the efficiency curve. We find a difference of 
0.1\% and take it as the systematic uncertainty due to the efficiency curve. 
We estimate the systematic uncertainties due to the line-shapes of the backgrounds by using the shapes from alternative ones. The line shape of one background component is changed each time, and the largest difference is used for the uncertainty. We change the relative ratio $f_{\rm FSR}$ by $1\sigma$ to evaluate the systematic 
uncertainty due to FSR. We change the Novosibirsk function to a Gaussian function to estimate the uncertainty due to the 
shape of the $\piz K\bar{K}\pi$ backgrounds, and change the shape from the inclusive MC sample to the Argus function~\cite{Argus} to estimate the uncertainty due to the shape of other backgrounds. 
We assume isospin symmetry between $\kkp$ and $\kskp$ channels in the nominal fit, and take the difference to the result without isospin conservation as the uncertainty associated with the isospin conservation assumption.
The systematic uncertainty associated with the damping function is estimated in two ways. One is varying the parameter $m_{\etacp}$ by $1\sigma$ in KEDR's damping function~\cite{KEDR}. The other one is modifying the function from the default KEDR's formula~\cite{KEDR} to the alternative CLEO's $\exp(-E^{2}_{\gamma}/8\beta^{2})$~\cite{CLEO}, where $\beta=0.033$. The maximum differences from the nominal results are taken as the relevant systematic uncertainties.
 
The uncertainty due to the number of $\psi(3686)$ events in the data sample is determined with inclusive hadronic 
$\psp$ decays. The 0.5\% is determined to be the systematic uncertainty associated with the number of $\psp$ events~\cite{Bes:data2021}.

\begin{table*}[htbp]
\centering
\footnotesize
\setlength{\tabcolsep}{5pt}
\renewcommand{\arraystretch}{1.2}
\caption{A comparison of the experimental measurements in this work with previous BESIII values and world average values for 
the $\eta_c(2S)$ mass and width, the BF $\BR(\ptge)$, and the partial width $\Gamma(\ptge)$.}
\begin{tabular}{c c c c c}
\hline\hline 
     &Mass~($\mevcs$)       & Width~($\mev$)  &$\BR(\ptge)~(\times10^{-4})$  &$\Gamma(\ptge)~(\kev)$\\
\hline
This work  & $3637.8 \pm 0.8 \pm 0.2$   &$10.5 \pm 1.7 \pm 3.5$ &$5.2 \pm 0.3 \pm 0.5~^{+~1.9}_{-~1.4}$ &$0.15~^{+~0.06}_{-~0.04}$\\
 BESIII~(2012)  &$3637.6 \pm 2.9 \pm 1.6$     &$16.9 \pm 6.4 \pm 4.8$ &$6.8 \pm 1.1 \pm 4.5$         &$0.20\pm 0.14$\\
World average &$3637.6 \pm 1.2$  &$11.3_{~-2.9}^{~+3.2}$ &$7 \pm 5$  & $0.21\pm 0.15$\\
\hline \hline
\end{tabular}
\label{Tab:BB_comp}
\end{table*}

The uncertainty due to the continuum process includes the number of events and the line shape. We vary the number of continuum events assuming it satisfies a Possion distribution, and take the difference to the nominal fit result as the corresponding uncertainty. We also vary the smooth parameter of RooKeysPdf~\cite{keyspdf}, which is extracted from the continuum line shape, from 1 to 6, and take the largest difference to the nominal fit result as the corresponding uncertainty.

We summarize the systematic uncertainties in Table~\ref{Tab:fit_sysunc}. We assume that all sources of systematic 
uncertainties are independent and combine them in quadrature to obtain the overall systematic uncertainty.

\section{Result and discussion}

Using a sample of $(27.08 \pm 0.14 ) \times 10^8~\psp$ decays collected by the BESIII detector, we measure the resonant parameters of 
the $\etacp$ and the product branching fraction $\BR(\ptge)\times \BR(\etacp \to K\bar{K}\pi)$ through the hadronic $\etacp$ decays $\kskp$ and $\kkp$ with improved precision. We measure the mass and width of the $\etacp$ to be 
$(3637.8 \pm 0.8 \pm 0.2)~\mevcs$ and $(10.5 \pm 1.7 \pm 3.5)~\mev$, respectively, and $\BR(\ptge) \times \BR(\etacp \to K\bar{K}\pi) = (0.97 \pm 0.06 \pm 0.09) \times$ $10^{-5}$, in which isospin symmetry is assumed. Combining our result 
with $\BR[(\etacp \to K\bar{K}\pi)] = (1.86^{+0.68}_{-0.49})\%$~\cite{whp}, we obtain $\BR(\ptge = 
(5.2 \pm 0.3 \pm 0.5^{+1.9}_{-1.4}) \times 10^{-4}$, where the last uncertainty is from the quoted $\BR(\etacp \to K\bar{K}\pi)$. The partial width $\Gamma(\ptge)$ is also determined to be $0.15^{+0.06}_{-0.04}~\kev$ by using the total width of the $\psp$~\cite{PDG}.

Compared with previous measurements~\cite{BES:2012uhz, PDG}, listed in Table~\ref{Tab:BB_comp}, our results for the mass and width of the $\etacp$ and $\BR(\ptge)$ are of comparable precision to the world average values. There is an improvement compared to the previous BESIII result reported in 2012\cite{BES:2012uhz}, while the last systematic uncertainty is not significantly reduced due to the large uncertainty of the quoted BF of $\etacp \to K\bar{K}\pi$, that is changed from $1.9\pm1.2\%$~\cite{BaBar} to $1.86^{+0.68}_{-0.49}\%$~\cite{whp}. Compared with the theoretical calculations ~\cite{Barnes:2005pb,Li:2007xr,Peng:2012tr} listed in Table~\ref{Tab:Width_theoretical}, our measurement of the partial width $\Gamma(\ptge)$ is consistent with all the theoretical predictions~\cite{Barnes:2005pb,Li:2007xr,Peng:2012tr} within two standard deviations. Therefore, to distinguish various theoretical models, the precision of the experimental measurement needs further improvement. It will be achieved by combining our results to an updated $\BR(\etacp \to K\bar{K}\pi)$ that may be obtained via two-photon fusion or $B$ decay processes in the $B$-factories.

\begin{acknowledgments}

The BESIII Collaboration thanks the staff of BEPCII and the IHEP computing center for their strong support. This work is supported in part by National Key R\&D Program of China under Contracts Nos. 2020YFA0406301, 2020YFA0406300, 2020YFA0406400; National Natural Science Foundation of China (NSFC) under Contracts Nos. 12275058, 11635010, 11735014, 11835012, 11935015, 11935016, 11935018, 11961141012, 12022510, 12025502, 12035009, 12035013, 12061131003, 12192260, 12192261, 12192262, 12192263, 12192264, 12192265, 12221005, 12225509, 12235017; the Chinese Academy of Sciences (CAS) Large-Scale Scientific Facility Program; the CAS Center for Excellence in Particle Physics (CCEPP); Joint Large-Scale Scientific Facility Funds of the NSFC and CAS under Contract No. U1832207; CAS Key Research Program of Frontier Sciences under Contracts Nos. QYZDJ-SSW-SLH003, QYZDJ-SSW-SLH040; 100 Talents Program of CAS; The Institute of Nuclear and Particle Physics (INPAC) and Shanghai Key Laboratory for Particle Physics and Cosmology; European Union's Horizon 2020 research and innovation programme under Marie Sklodowska-Curie grant agreement under Contract No. 894790; German Research Foundation DFG under Contracts Nos. 455635585, Collaborative Research Center CRC 1044, FOR5327, GRK 2149; Istituto Nazionale di Fisica Nucleare, Italy; Ministry of Development of Turkey under Contract No. DPT2006K-120470; National Research Foundation of Korea under Contract No. NRF-2022R1A2C1092335; National Science and Technology fund of Mongolia; National Science Research and Innovation Fund (NSRF) via the Program Management Unit for Human Resources \& Institutional Development, Research and Innovation of Thailand under Contract No. B16F640076; Polish National Science Centre under Contract No. 2019/35/O/ST2/02907; The Swedish Research Council; U. S. Department of Energy under Contract No. DE-FG02-05ER41374.

\end{acknowledgments}

\bibliography{draft}

\end{document}